\documentclass[showpacs,prd,10pt,twocolumn,superscriptaddress,%
nofootinbib,aps]{revtex4-2}

\usepackage{ae}
\usepackage{bm} 
\usepackage{xcolor}
\usepackage{amsmath}
\usepackage{amssymb}
\usepackage{bm}
\usepackage{amsfonts}
\usepackage{graphicx}
\usepackage{slashed}
\usepackage{wasysym}
\usepackage{units}
\usepackage{hyperref}
\usepackage{feynmp-auto}
\usepackage{tikz}
\usepackage[compat=1.1.0]{tikz-feynman}
\usetikzlibrary{external}
\usepackage{caption}
\usepackage{subcaption}
\usepackage{graphicx}

\newcommand{\Nf}{N_{\text{f}}}

\usepackage[utf8]{inputenc}

\makeatletter
\def\fps@figure{ht}
\def\fps@table{ht}
\makeatother

\allowdisplaybreaks

\definecolor{shiraz}{rgb}{0.6,0,0.4}

\begin{document}

\setlength{\unitlength}{1mm}

\title{Fermionic Fixed-Point Structure of Asymptotically Safe QED with a 
Pauli Term}

\author{Holger Gies}
\affiliation{Theoretisch-Physikalisches Institut, Abbe Center of Photonics, 
Friedrich-Schiller-Universit\"at Jena, Max-Wien-Platz 1, 07743 Jena, Germany}
\affiliation{Helmholtz-Institut Jena, Fr\"obelstieg 3, 07743 Jena, Germany}
\affiliation{GSI Helmholtzzentrum für Schwerionenforschung, Planckstr. 1, 
64291 Darmstadt, Germany} 
\author{Kevin K. K. Tam}
\affiliation{Theoretisch-Physikalisches Institut, Abbe Center of Photonics, 
Friedrich-Schiller-Universit\"at Jena, Max-Wien-Platz 1, 07743 Jena, Germany}

\date{\today}

\begin{abstract} 
We test the physical viability of a recent proposal for an 
asymptotically safe modification of quantum electrodynamics (QED), whose 
ultraviolet physics is dominated by a non-perturbative Pauli spin-field 
coupling. We focus in particular on its compatibility with the absence of 
dynamical generation of fermion mass in QED. Studying the renormalization group 
flow of chiral four-fermion operators and their fixed points, we 
discover a distinct class of behavior compared to the standard picture of 
fixed-point annihilation at large gauge couplings and the ensuing formation of 
chiral condensates. Instead, transcritical bifurcations, where the fixed points 
merely exchange infrared stability, are observed. Provided that 
non-chiral operators remain irrelevant, our theory accommodates a universality 
class of light fermions for $\Nf > 1$ irreducible Dirac flavors. 
On the contrary, in the special case of $\Nf = 1$ flavor, this comes only at the 
expense of introducing one additional relevant parameter. 
\end{abstract}

\maketitle

\section{Introduction}
\label{sec:intro}

Quantum electrodynamics (QED) is an extremely well-tested theory, 
exhibiting remarkable agreement with precision experiments at low energies 
\cite{QEDtest_PhysRevLett.130.071801,QEDtest_Ali:1988ts,
QEDtest_10.2307/j.ctt2jc8td,QEDtest_Gabrielse:2009tps,Fan:2022eto}. Of course, 
high-energy tests are also passed by the theory though at lower precision 
\cite{ATLAS:2017fur,CMS:2018erd} and ultimately require the embedding of QED 
into the  
electroweak sector of the standard model. Still, the high-energy behavior 
of pure QED remains of interest in its own right, as it has constituted a 
puzzle since the early days of quantum field theory: perturbation theory 
predicts a divergence of the minimal gauge coupling at a finite Landau pole 
\cite{LandauPole_osti_4412940,Landau:1955}. While this may simply signal the 
expected breakdown of perturbation theory in the strong-coupling regime, the 
conclusion of the existence of a finite scale of maximum ultraviolet (UV) 
extension is supported by lattice simulations 
\cite{Lattice_Kim_2001,Lattice_PhysRevD.65.054015,Lattice_PhysRevLett.80.4119} 
and functional methods \cite{Functional_PhysRevLett.93.110405}. 

The picture obtained from such nonperturbative methods is, however, decisively 
different from simple perturbation theory: a strong coupling phase of QED -- 
even if it existed -- can generically not be connected by a line of constant 
physics to physical QED because of chiral symmetry breaking. Strong gauge 
interactions induce fermion mass generation with masses on the order of the 
high scale being incompatible with the observed existence of a light electron. 
In continuum computations, the symmetry breaking can be traced back to 
fermionic self-interactions turning into relevant operators at strong coupling 
and triggering condensate formation 
\cite{chiSB,chiSB_Aoki_1997,Functional_PhysRevLett.93.110405}.
The corresponding long-range limit of such a theory would then be a free photon 
theory. 

As a resolution, a recent proposal has been based on the observation that the 
Pauli spin-field coupling term $\bar\psi \sigma_{\mu\nu}F^{\mu\nu}\psi$ has the 
potential to screen the Landau pole -- 
and thus the strong-coupling regime -- within an effective field theory 
\cite{Djukanovic:2017thn}. In fact, a self-consistent analysis of pure QED with 
a Pauli term has provided evidence for the existence of 
interacting fixed-points potentially rendering QED 
asymptotically safe \cite{Pauli_Gies_2020, 
Pauli_https://doi.org/10.48550/arxiv.2210.11927} and thus high-energy complete. 
As a dimension-5 operator with only a single derivative with respect to the 
photon, the Pauli term represents the unique next-to-leading-order contribution 
in a combined derivative and power-counting operator expansion of the effective 
action. 

By the techniques of functional renormalization, the extended theory 
space has been shown to include two non-trivial fixed points $\mathcal{B}$ and 
$\mathcal{C}$ at vanishing gauge coupling \cite{Pauli_Gies_2020}, each of which 
provides an ultraviolet (UV) completion of QED as an asymptotic safety scenario. 
Specifically, the fixed point $\mathcal{C}$ occurring at a finite Pauli coupling 
$\kappa$ is compatible with a renormalization 
group (RG) trajectory reproducing the long-range values of phenomenological 
QED. As it features three relevant directions, the long-range physics is fully 
predictive, once three parameters have been fixed by experiment (e.g., the 
electron mass, the fine structure constant, and the anomalous magnetic moment 
of the electron). 
(By contrast, fixed point $\mathcal{B}$ predicts unphysically large values of 
the anomalous magnetic moment in the infrared (IR); while being potentially 
consistent and UV complete, this universality class is observationally not 
viable.) 

In view of the impossibility to connect conventional strong-coupling QED with 
the observed existence of light electrons, an obvious question needs to be 
answered: does an asymptotically safe UV completion based on the Pauli term 
preserve chiral symmetry along its RG trajectories towards the infrared (IR)?
This is not at all evident, since fixed point $\mathcal{C}$ -- though 
featuring a vanishing mass -- occurs at a deeply nonperturbative value of the 
Pauli coupling $\kappa^* = 3.82$, independently of fermion flavor number 
\cite{Pauli_https://doi.org/10.48550/arxiv.2210.11927}. 

To further scrutinize the physical relevance of this continuum theory, we go 
beyond the Pauli term in the truncation of the effective action. Operators of 
particular interest are given by dimension-6 four-fermion channels of the 
Nambu-Jona-Lasinio (NJL) type, which appear in an effective theory of 
spontaneous chiral symmetry breaking in quantum chromodynamics 
\cite{NJL_PhysRev.122.345, NJL_PhysRev.124.246}. Just as the formation of 
chiral 
condensates is responsible for the constituent quark masses, the focus of this 
work is to investigate whether the strong coupling regime at fixed point 
$\mathcal{C}$ dynamically generates mass at a UV scale, for example at the 
Planck scale, which would be in contradiction to the observation of light 
fermions of the Standard Model. Similar problems are known to impede non-trivial 
formulations of  
pure QED \cite{Lattice_Kim_2001,Functional_PhysRevLett.93.110405, chiSB, %
chiSB_Aoki_1997}. 

In QED, chiral symmetry is broken explicitly by the mass term. In the same 
manner, the Pauli spin-field coupling is also a source of explicit breaking, 
both of which we consider as small in agreement with observation. While such 
small breakings allow for the appearance of many further four-fermion 
interactions, we concentrate here on an otherwise 
Fierz-complete basis of NJL-type channels, assuming that they play a dominant 
role in the case of interaction-induced dynamical chiral symmetry breaking. 
This assumption is similar to low-energy effective theories for QCD where 
explicit chiral breaking terms can be treated as a small perturbation. 

In this setting, we discover that the distinct coupling of 
the two NJL-type channels by the Pauli term qualitatively alters the 
bifurcation behavior known from strong QED or QCD: instead 
of annihilation upon collision, the NJL fixed points merely exchange stability 
such that an IR attractor persists at arbitrarily strong Pauli coupling for more 
than $\Nf = 1$ fermion flavor. In such cases, there exists a universality class 
where the RG flow remains bounded and mass generation can be avoided without 
further fine-tuning. A similar conclusion can be drawn from our initial analysis 
for fixed point $\mathcal{B}$, despite the different role played by the 
aforementioned bifurcation. 

The structure of this paper is as follows: In Sect. \ref{sec:model}, we 
introduce the abelian gauged NJL model with a Pauli term. Sect. 
\ref{sec:floweq} then presents the corresponding RG flow equation. Sect. 
\ref{sec:fixedpoint} is allocated to analyzing the fixed point structure in the 
four-fermion sector of our theory, drawing comparisons to previous results in 
relation to the pure NJL model (\ref{subsec:PureNJL}) and the introduction of a 
gauge field (\ref{subsec:gaugecoupling}). 

\section{Gauged NJL Model with Pauli Term}
\label{sec:model}

We consider the massless limit of an abelian gauged NJL model with a Pauli 
term. Satisfying Osterwalder-Schrader reflection positivity in Euclidean 
spacetime, the effective action reads 
\begin{align}
    \Gamma_k
    = \int d^4x \, \Bigl\{&\bar{\psi}^a \left(iZ_\psi \slashed{\partial} + 
\bar{e} \slashed{A} + i \bar{\kappa} \sigma_{\mu \nu} F^{\mu \nu}\right) \psi^a 
\nonumber \\ 
    &+ \frac{Z_A}{4} \, F_{\mu\nu} F^{\mu\nu} + \frac{Z_A}{2\xi} 
\left(\partial_\mu A^\mu\right)^2 \nonumber \\ 
    &+ \frac{1}{2}\bar{\lambda}_+ \left(\text{V} + \text{A} \right) + 
\frac{1}{2}\bar{\lambda}_- \left(\text{V} - \text{A} \right) \Bigr\}. 
    \label{eq:GammaK}
\end{align}
Here $a = 1, \dots, \Nf$ labels the Dirac flavors $\psi^a$ interacting with a 
U(1) gauge field $A_ 
\mu$. The couplings $\bar{e}$, $\bar{\kappa}$ and $\bar{\lambda}_\pm$, as well 
as the wave function renormalizations $Z_{\psi,A}$, are dependent on the RG 
scale $k$. We further work in the Landau gauge $\xi = 0$ as a fixed point of the 
RG flow \cite{LandauGauge_Litim_1998, 
LandauGauge_https://doi.org/10.48550/arxiv.hep-th/9506019}. For our purposes, it 
suffices to consider the point-like approximation where the four-fermion 
couplings $\bar{\lambda}(p_1, p_2, p_3) \to \bar{\lambda}(0, 0, 0)$ are 
approximated by their low-momentum limit \cite{4Fermi_Braun_2012}. Neglecting 
the explicit breaking of a chiral SU$(\Nf)_\text{L} \otimes$ SU$(\Nf)_\text{R}$ 
symmetry by the Pauli term, the four-fermion channels 
\begin{equation}
    \left(\text{V} \pm \text{A} \right) \equiv \left(\bar{\psi}^a \gamma_\mu 
\psi^a\right)^2 \pm \left(\bar{\psi}^a i\gamma_\mu \gamma_5 \psi^a\right)^2, 
    \label{eq:DefVA}
\end{equation}
would form a Fierz-complete basis under the RG flow. The $\left(\text{V} 
+ \text{A} \right)$ channel is Fierz equivalent to the conventional NJL 
channel. In the limit of vanishing $\bar{\kappa},\bar{\lambda}_\pm$ (and upon 
inclusion of an explicit fermion mass term), the present model is identical to 
QED. If RG trajectories exist that match the QED long-range behavior, then a 
high-energy complete trajectory in the present model can be viewed as a 
UV-complete version of QED. In the search for scale-invariant fixed points 
facilitating UV-complete trajectories, it is convenient to define 
further dimensionless renormalized couplings 
\begin{equation}
    \lambda_\pm = \frac{k^2 \bar{\lambda}_\pm}{Z_\psi^{2}} , \quad e = 
\frac{\bar{e}}{Z_\psi \sqrt{Z_A}}, \quad \kappa = \frac{k \bar{\kappa}}{Z_\psi 
\sqrt{Z_A}}. 
    \label{eq:dimrencoup}
\end{equation}
For the present study, we use the functional RG based on the effective average 
action $\Gamma_k$ which interpolates between the classical bare action 
$\Gamma_{k\rightarrow \Lambda} = S$ and the full quantum effective action 
$\Gamma_{k\rightarrow 0} = \Gamma$ \cite{FRG_BERGES2002223, FRG_Gies_2012, 
FRG_Nagy_2014, FRG_Delamotte_2012, FRG_PAWLOWSKI20072831, 4Fermi_Braun_2012}. 
Defining the RG time as $t = \ln k$, the flow through theory space is governed 
by the Wetterich equation 
\cite{Wetterich:1992yh,Bonini:1992vh,Ellwanger:1993mw,Morris:1993qb}
\begin{equation}
    \partial_t \Gamma_k = \frac{1}{2} \text{STr}\left[\partial_t R_k 
\left(\Gamma^{(2)}_k[\phi] + R_k \right)^{-1}\right],  
    \label{eq:wetterich}
\end{equation}
where $R_k(p^2)$ acts as a momentum-dependent regulator, screening the 
contribution of IR modes with momenta below the cutoff $k$. 

\section{Flow Equations}
\label{sec:floweq}

\begin{figure}[ht!]
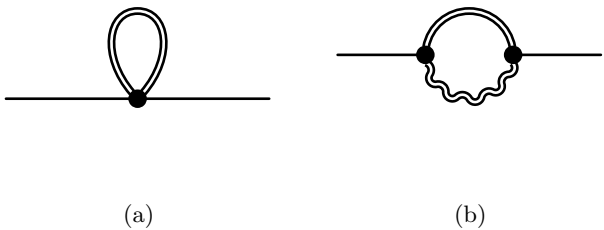

    \begin{center}
     \begin{subfigure}[ht!]{.49\linewidth}
        \centering
        \includegraphics{figures/f3b.1}
        \caption{}
        \label{fig:tadpole}
     \end{subfigure}
     \hfill
     \begin{subfigure}[ht!]{.49\linewidth}
         \centering
        \includegraphics{figures/loopy.1}
        \caption{}
        \label{fig:selfenergy}
     \end{subfigure}
     \end{center}
\caption{1PI Feynman diagrams contributing to the fermionic anomalous 
dimension $\eta_\psi$. a) The tadpole diagrams vanish in the point-like limit 
where the momentum dependence of the NJL couplings $\lambda_\pm$ is neglected. 
b) The self-energy diagrams result in the value of $\eta_\psi = -1$ at fixed 
point $\mathcal{C}$, which renders both Pauli coupling $\kappa$ and NJL coupling 
$\lambda_\pm$ perturbatively marginal in the absence of higher-order terms.}   
        \label{fig:anomdim}
\end{figure}

With the effective action $\Gamma_k$ expressed in terms of the 
operators in \eqref{eq:GammaK}, the Wetterich equation \eqref{eq:wetterich} can 
be projected onto the four-fermion sector to yield the beta functions 
\begin{widetext}
    \begin{align}
    \partial_t \lambda_+ 
    &= (2+2\eta_\psi) \, \lambda_+ + 4v_4 \, l^{(\text{F})}_4(0,0) \left[6\lambda_+^2 +\left(d_\gamma N_f +4\right)\lambda_+ \lambda_-\right] \label{eq:LamPlusBeta} \\
    &\quad -24 v_4 \, l^{(\text{B}, \, \text{F})}_4(0, 0) \, e^2 \lambda_+ - 48 v_4 \, l^{(1, \, \text{B}, \, \text{F})}_4(0, 0) \, \kappa^2 \lambda_- \nonumber \\
    & \quad+ 6 v_4 \, l^{(\text{B}^2, \, \text{F})}_4(0, 0) \, e^4 - 48 v_4 \, l^{(1, \, \text{B}^2, \, \text{F})}_4(0, 0) \, e^2 \kappa^2 + 96 v_4 \, l^{(2, \, \text{B}^2, \, \text{F})}_4(0, 0) \, \kappa^4, \nonumber \\ \nonumber \\
    \partial_t \lambda_- 
    &= (2+2\eta_\psi)\lambda_- + 2v_4 \, l^{(\text{F})}_4(0,0) \left[\left(d_\gamma N_f -4\right)\lambda_-^2 +d_\gamma N_f \lambda_+^2\right] \label{eq:LamMinusBeta} \\
    &+ 24 v_4 \, l^{(\text{B}, \, \text{F})}_4(0, 0) \, e^2 \lambda_- - 48 v_4 \, l^{(1, \, \text{B}, \, \text{F})}_4(0, 0) \, \kappa^2 \lambda_+ \nonumber \\
    &- 6 v_4 \, l^{(\text{B}^2, \, \text{F})}_4(0, 0) \, e^4 + 48 v_4 \, l^{(1, \, \text{B}^2, \, \text{F})}_4(0, 0) \, e^2 \kappa^2 - 96 v_4 \, l^{(2, \, \text{B}^2, \, \text{F})}_4(0, 0) \, \kappa^4. \nonumber
\end{align}
\end{widetext}
Here we have adopted the notation for threshold functions introduced in 
\cite{Pauli_Gies_2020}. The anomalous dimension 
\begin{equation}
\eta_\psi=- \partial_t \ln
  Z_\psi \label{eq:eta}
\end{equation}
implements an RG improvement by resummation of 1PI diagrams contributing to the 
propagator, as depicted in Figure \ref{fig:anomdim}. In the point-like limit, 
four-fermion corrections to $\eta_\psi$ must vanish, as momentum conservation in 
the tadpole diagram of Figure \ref{fig:tadpole} ensures the independence of the 
loop momentum from external momentum. As such, the fermionic anomalous dimension 
at fixed point $\mathcal{C}$ remains at the value of $\eta_\psi^* = -1$ due to 
Pauli contributions (Figure \ref{fig:selfenergy}). This precisely renders the 
dimension-5 Pauli operator marginally relevant in $d=4$, even before considering 
higher-order diagrams.  

Likewise, the scaling terms in 
Eqs.~\eqref{eq:LamPlusBeta}-\eqref{eq:LamMinusBeta} vanish, and thus the 
relevance of the NJL channels is decided entirely by the higher-order terms of 
the beta functions, as represented by Figure \ref{fig:4fermi}. 

\section{Fixed Point Analysis}
\label{sec:fixedpoint}

\subsection{Pure NJL Model}
\label{subsec:PureNJL}

\begin{figure}
    \centering
    \includegraphics[width=\linewidth]{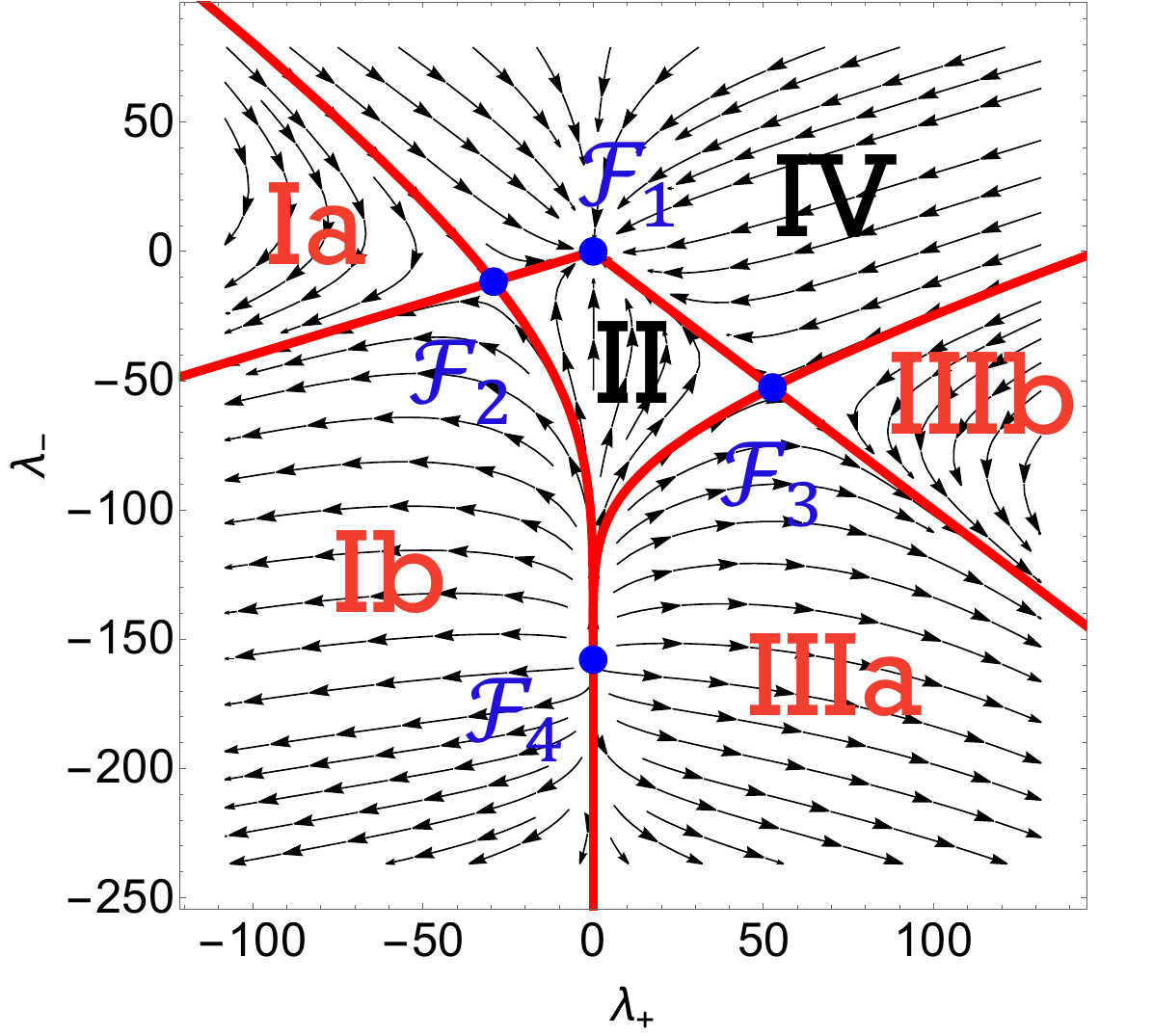}
    \caption{Phase diagram of the NJL theory subspace spanned by the 
$\left(\text{V} \pm \text{A} \right)$ channels for $\Nf = 2$ irreducible Dirac 
flavors. Arrows indicate flow towards the infrared. Separatrices (red curves) 
flowing between fixed points $\mathcal{F}_i$ (blue points)    delineate a 
universality class of light fermions (regions II \& IV; black) as observed in 
nature. On the contrary, the RG flow in regions I \& III diverge, heralding the 
onset of dynamical mass generation. For $\Nf = 1$, $\mathcal{F}_4$ lies at 
infinity.} 
    \label{fig:PureNJL}
\end{figure}

Before we examine the fate of the NJL channels in the presence of the Pauli 
term, we first review the RG flow of the pure NJL model ($e = \kappa = \eta_\psi 
= 0$) \cite{Gies:2003dp,4Fermi_Braun_2012,2Chan_Eichhorn_2011}, shown in Figure 
\ref{fig:PureNJL} for the case of $\Nf > 1$ irreducible flavors, for which the 
Dirac representation has dimension $d_\gamma = 4$. As the beta 
functions~\eqref{eq:LamPlusBeta}-\eqref{eq:LamMinusBeta} form a pair of 
quadratic functions of the NJL couplings $\lambda_\pm$, there exist in general 
four fixed points $\mathcal{F}_i = \left(\lambda_+^{*i}, \lambda_-^{*i}\right), 
\, i = 1,\dots,4$. We quantify the fixed-point properties in terms of 
their critical exponents $\theta$ which are related to the eigenvalues of the 
stability matrix $B_{ij}:= \partial (\partial_t \lambda_i)/\partial 
\lambda_j|_{\lambda^*} $: $\theta = - \text{eig} (B)$. Positive values of 
$\theta$ denote RG relevant directions that correspond to physical parameters 
to be fixed. Negative exponents, in turn, characterize RG irrelevant directions 
that do not exert an influence on the long-range IR physics. 

As expected from power counting, the Gaussian fixed point 
$\mathcal{F}_1$ is purely IR attractive with critical exponents both being 
$\theta=-2$. The interacting fixed points $\mathcal{F}_2$ and $\mathcal{F}_3$ 
each has one relevant direction, while the fourth fixed point $\mathcal{F}_4$ 
is purely IR repulsive, i.e. relevant. Each of the fixed points 
$\mathcal{F}_{2,3,4}$ has one relevant eigendirection ($\theta=2$) pointing 
along the line that connects the fixed point $\mathcal{F}_{i\geq 2}$ 
with the Gaussian fixed point $\mathcal{F}_1$. 
This is in line with general theorems \cite{Gies:2003dp,Gehring:2015vja}. It 
is straightforward to also compute the remaining critical exponents 
analytically.

\begin{figure}[ht!]
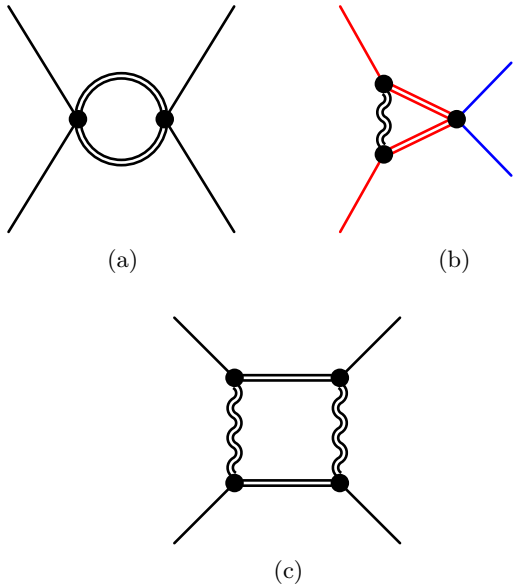

\begin{center}
\begin{subfigure}{.49\linewidth}
    \centering
    \includegraphics{figures/f4c.1}
    \caption{}
    \label{fig:fermionloop}
\end{subfigure}
    \hfill
\begin{subfigure}{.49\linewidth}
    \centering
    \includegraphics{figures/f2p.1}
    \caption{}
    \label{fig:SingleExchange}
\end{subfigure}

\bigskip
\begin{subfigure}{\linewidth}
    \centering
    \includegraphics{figures/feyngraph1.1}
    \caption{}
    \label{fig:DoubleExchange}
\end{subfigure}
\end{center}
\caption{1PI Feynman diagrams contributing to the flow of the four-fermion 
vertex. a) The fermionic loops carry flavor number dependence such that the 
$\lambda_-^2$ contribution vanishes for a single irreducible flavor. b) The 
triangular diagrams only contribute to the RG flow in the chirally invariant NJL 
subspace when the photon is exchanged between fermions of identical flavor. 
Moreover, the flows of $\lambda_\pm$ are maximally coupled by these diagrams, 
unlike their gauge coupling counterparts. c) The box diagrams can induce a 
finite NJL coupling $\lambda_\pm$ purely through photonic fluctuations.} 
\label{fig:4fermi}
\end{figure}

We observe that the purely IR-repulsive $\mathcal{F}_4$ moves towards 
infinity for $\Nf \to 1$ flavor. This is because the beta function 
\eqref{eq:LamMinusBeta} becomes linear in $\lambda_-$, with the vanishing 
fermionic loop contribution $\sim \lambda_-^2$ of Figure \ref{fig:fermionloop}. 

The universality class defined by the Gaussian fixed point $\mathcal{F}_1$ 
corresponds to a chirally symmetric phase (regions II \& IV) with massless 
fermions, as the NJL couplings $\lambda_\pm$ remain finite under the RG flow 
and approach zero in the long-range limit. Meanwhile, initial conditions within 
regions I \& III lead to divergence at a finite RG scale $k_{SB}$, signaling 
the formation of a condensate which dresses the fermions with a mass $m_\psi 
\sim k_{SB}$. This phase of the model is used in low-energy QCD effective 
models. 

In the simplest incarnation of the NJL model, the coupling $\lambda_-$ is 
set to zero, and the coupling $\lambda_+$ corresponds to $(-2)$ times the more 
familiar scalar-pseudoscalar channel $(\text{S}-\text{P})$. The fixed-point 
$\mathcal{F}_2$ (projected on the $\lambda_+$ axis) then defines the 
NJL critical coupling that separates the chirally symmetric weak-coupling phase 
from the chirally broken phase at strong coupling.

\subsection{Finite Gauge Coupling}
\label{subsec:gaugecoupling}

With a nonzero gauge coupling $e$, the beta 
functions~\eqref{eq:LamPlusBeta}-\eqref{eq:LamMinusBeta} reproduce the known 
result where the Gaussian fixed point $\mathcal{F}_1$ is annihilated by 
collision with $\mathcal{F}_2$ at a critical value $e_{\text{crit}}$. This 
-- in a nutshell -- illustrates the relevant mechanism that screens 
the perturbative Landau pole and inhibits a UV completion of long-range QED: 
even if QED were UV complete in the strong coupling region, it would exhibit 
high-scale chiral symmetry breaking and mass generation in contradiction to 
the observed light mass of the electron 
\cite{chiSB,Lattice_PhysRevLett.80.4119,Lattice_PhysRevD.65.054015,%
Functional_PhysRevLett.93.110405}. In analogous nonabelian settings, the 
similar mechanism involving the strong gauge coupling triggers the dynamical 
mass generation in the IR limit of quantum chromodynamics 
\cite{4Fermi_Braun_2012, FPAnnihilation_article, 
FPAnnihilation_Braun_2006,FPAnnihilation_cite-key,FPAnnihilation_Gies_2006,%
Mitter:2014wpa,Braun:2014ata,Eichmann:2016yit,Binosi:2016wcx,Cyrol:2017ewj}. 

Such an effect is already captured by the Fierz-incomplete 
single-channel approximation $\lambda_- \equiv 0$. The remaining beta function, 
now represented by a parabola, is shifted vertically by the finite gauge 
coupling until the fixed points undergo a saddle-node bifurcation into the 
complex plane. 

\subsection{Finite Pauli Coupling}

In contrast to the theory space spanned by minimally coupled QED, the 
inclusion of the Pauli coupling has provided evidence for the existence of a 
new universality class governed by a non-Gaussian fixed point, called fixed 
point $\mathcal{C}$ in \cite{Pauli_Gies_2020}. This fixed point occurs at 
$e^*=0$, $\kappa^*\simeq 3.82$ and $\eta_\psi=-1$ with the Pauli coupling and 
the minimal coupling corresponding to relevant directions (in addition to the 
massive perturbation).

With regard to Eqs.~\eqref{eq:LamPlusBeta} and \eqref{eq:LamMinusBeta}, we note 
that the only qualitative difference  
lies in the terms $\partial_t \lambda_\pm \sim \kappa^2 \lambda_\mp$ 
corresponding to the exchange of a single photon (Figure 
\ref{fig:SingleExchange}). Unlike their gauge coupling counterparts, these terms 
are non-diagonal in the $\lambda_\pm$ basis. Ultimately, this is due to the 
anticommutativity of all Dirac matrices with the $\gamma_5$ from the axial 
vector channel in Eq. \eqref{eq:DefVA}. 

\begin{figure}[ht!]
    \centering
    \includegraphics[width=\linewidth]{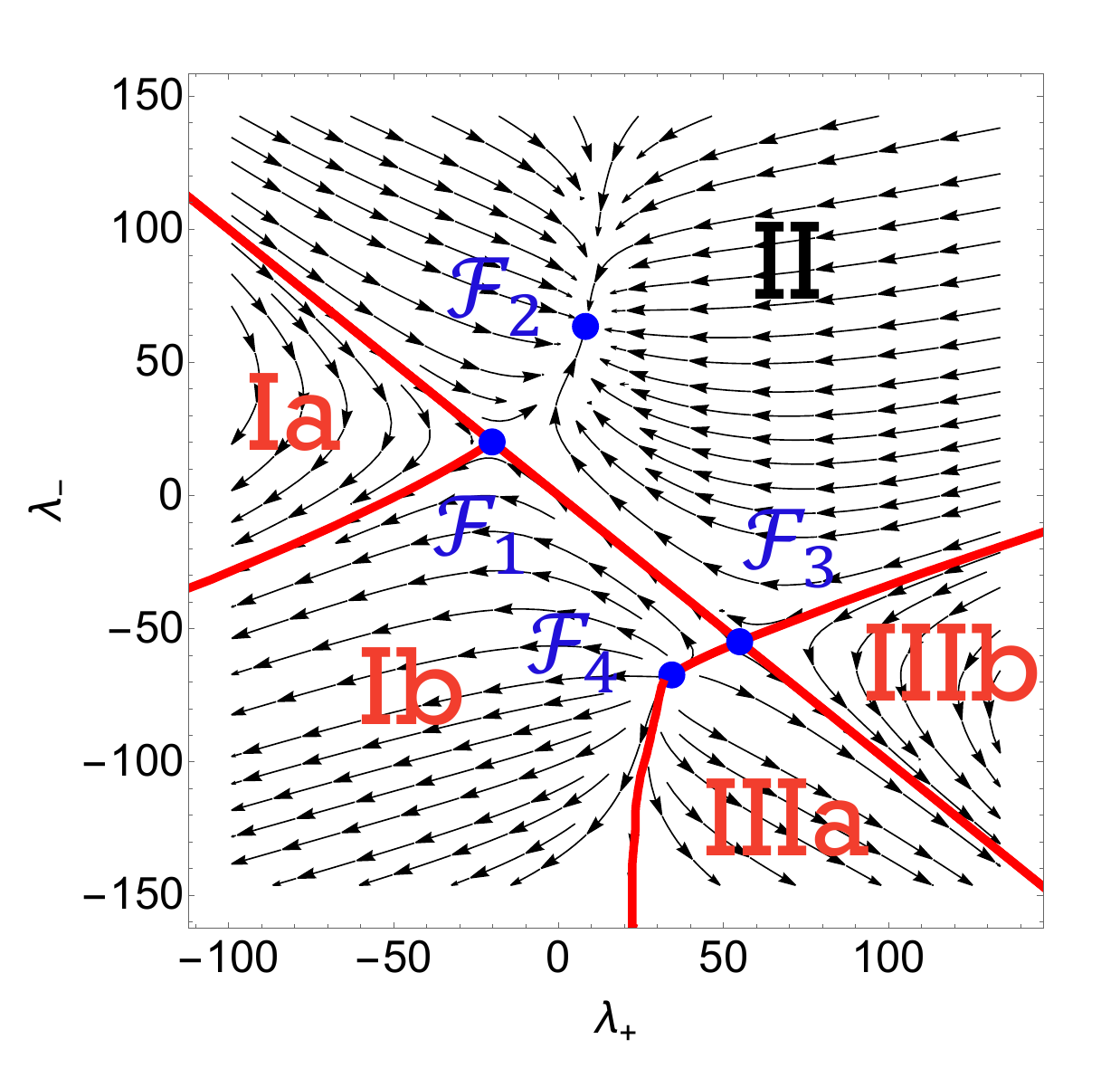}
    \caption{Phase diagram of the four-fermion subspace spanned by the NJL-type $\left(\text{V} \pm \text{A} \right)$ channels at the Pauli-induced 
fixed point $\mathcal{C}$ ($\kappa^* = 3.82, \eta_\psi = -1$) for $\Nf = 2$ 
irreducible Dirac flavors. Arrows indicate flow towards the infrared. Compared 
to Figure \ref{fig:PureNJL}, $\mathcal{F}_1$ and $\mathcal{F}_2$ have undergone 
a transcritical bifurcation and exchanged their stability. There remains a 
universality class of light fermions (region II).} 
    \label{fig:FPC_Nf2}
\end{figure}

For $\Nf>1$ we again observe a collision between 
fixed points $\mathcal{F}_1$ and $\mathcal{F}_2$ at a critical value 
$\kappa_\text{crit}$, but instead of annihilation, they merely exchange 
stability such that both fixed points continue to exist 
in the real coupling plane and $\mathcal{F}_2$ is now purely IR attractive. 
The result of this transcritical bifurcation is shown in Figure 
\ref{fig:FPC_Nf2}. The persistence of this attractor in the strong coupling 
regime maintains a universality class where mass generation is avoided (region 
II). This effect is not captured in the single-channel approximation 
$\lambda_- \equiv 0$. 

For an interpretation of the fixed-points as possible routes to UV 
completion or as critical couplings defining universality classes, we summarize 
their critical exponents in Tab.~\ref{tab:critexp}. With the gauge system being 
at fixed point $\mathcal{C}$ with a non-Gaussian Pauli coupling where 
$\eta_\psi=-1$, the naive non-Gaussian scaling of the four-fermion couplings 
suggests that the largest critical exponent should be close to zero. We 
therefore use large deviations from this expectation as an indication for 
sizable truncation artefacts. This reasoning is analogous to that applied to 
four-fermion models beyond $2$ dimensions  
\cite{Gies:2003dp,Gies:2010st,Janssen:2012pq,Gehring:2015vja,4Fermi_Braun_2012}
.

From this perspective, $\mathcal{F}_3$ and $\mathcal{F}_4$ represent 
fixed points with large deviations from the expected scaling that are likely to 
be dominated by truncation artefacts. While we do not expect them to persist in 
larger truncations and they should thus not be used for a construction of UV 
complete trajectories, they and their separatrices may still be used as a 
qualitative estimate of the boundaries of the chirally symmetric phase II.

\begin{table}[htbp]
	\begin{tabular}{c|c|c|c|c}
& $\mathcal{F}_1$ & $\mathcal{F}_2$ & $\mathcal{F}_3$ & $\mathcal{F}_4$
 \\ \hline
$\Nf=2$: & $(0.102,-4.41)$ & $(-0.101,-4.44)$ & $(4.41,-5.78)$ & $(10.0,2.75)$\\
\hline
$\Nf=1$: & $(3.14,-2.36)$ & -- & $(2.36,-6.30)$ & -- \\
 \end{tabular}
\caption{Critical exponents $\theta$ of the fixed points in the plane of 
four-fermion interactions with the gauge system being at the fixed point 
$\mathcal{C}$ with a non-Gaussian Pauli coupling. }
\label{tab:critexp}
\end{table}

By contrast, $\mathcal{F}_1$ and $\mathcal{F}_2$ exhibit small leading 
exponents close to zero. Fixed point $\mathcal{F}_2$ with two negative exponents 
is fully IR attractive and thus should be viewed as a \textit{shifted Gaussian 
fixed point} \cite{2Chan_Eichhorn_2011,Eichhorn:2012va}, playing the role of 
the Gaussian fixed point with a location at finite coupling due to the residual 
non-Gaussian interactions induced by the Pauli coupling. On the other hand, 
$\mathcal{F}_1$ also exhibits small deviations from the expected scaling with 
a relevant direction that points approximately along the NJL channel. Whether 
or not $\mathcal{F}_1$ could be used to define UV complete trajectories should 
be checked in future investigations. For the present work, we focus on the 
existence of $\mathcal{F}_2$ as a completely attractive fixed point.
This establishes that we 
find a qualitatively identical phase diagram to Figure \ref{fig:FPC_Nf2} for
more than one fermion flavor even for the case that the Pauli coupling
$\kappa^*$ is near fixed point $\mathcal{C}$. This statement holds
independently of flavor number $\Nf$
\cite{Pauli_https://doi.org/10.48550/arxiv.2210.11927} with the minor difference
that the magnitude of  $\kappa^*$ is insufficient to induce the collision
between $\mathcal{F}_1$ and $\mathcal{F}_2$ for larger flavor numbers > 5.25; in
such cases, $\mathcal{F}_1$ simply remains the shifted Gaussian fixed point.
A strong Pauli coupling phase of QED could thus allow for the 
construction of UV complete trajectories without being endangered by chiral 
symmetry breaking in contradistinction to a strong minimal coupling phase.
Incidentally, a further transcritical 
bifurcation occurs for $\Nf \gtrsim 4.94$ between $\mathcal{F}_3$ and 
$\mathcal{F}_4$. This, however, leaves our conclusions about UV 
completion in the symmetric phase unaffected.

In the special case of $\Nf=1$, as with the pure NJL model of Subsection 
\ref{subsec:PureNJL}, the quadratic term in $\lambda_-$ in equation 
\eqref{eq:LamMinusBeta} vanishes. Combined with the vanishing scaling term due 
to $\eta_\psi = -1$, all dependences on $\lambda_-$ drop out from the beta 
function. A transcritical bifurcation between $\mathcal{F}_1$ and 
$\mathcal{F}_2$ is still observed, but the purely attractive $\mathcal{F}_2$ 
then lies at infinity along with the purely repulsive $\mathcal{F}_4$. 
While this offers, in principle, a construction of a similar UV complete 
scenario as in the $\Nf>1$ case, the inherently large coupling values make it 
difficult to control the expansion scheme. The remaining two fixed points at 
finite coupling values $\mathcal{F}_1$ and $\mathcal{F}_3$ show large leading 
exponents with large deviations from the expected scaling, cf. 
Tab.~\ref{tab:critexp}. Our present study therefore does not allow us to draw any 
definite conclusions about the existence of UV complete trajectories controlled 
by the Pauli coupling for the special case of $\Nf=1$.

\begin{figure}[ht!]
    \centering
    \includegraphics[width=\linewidth]{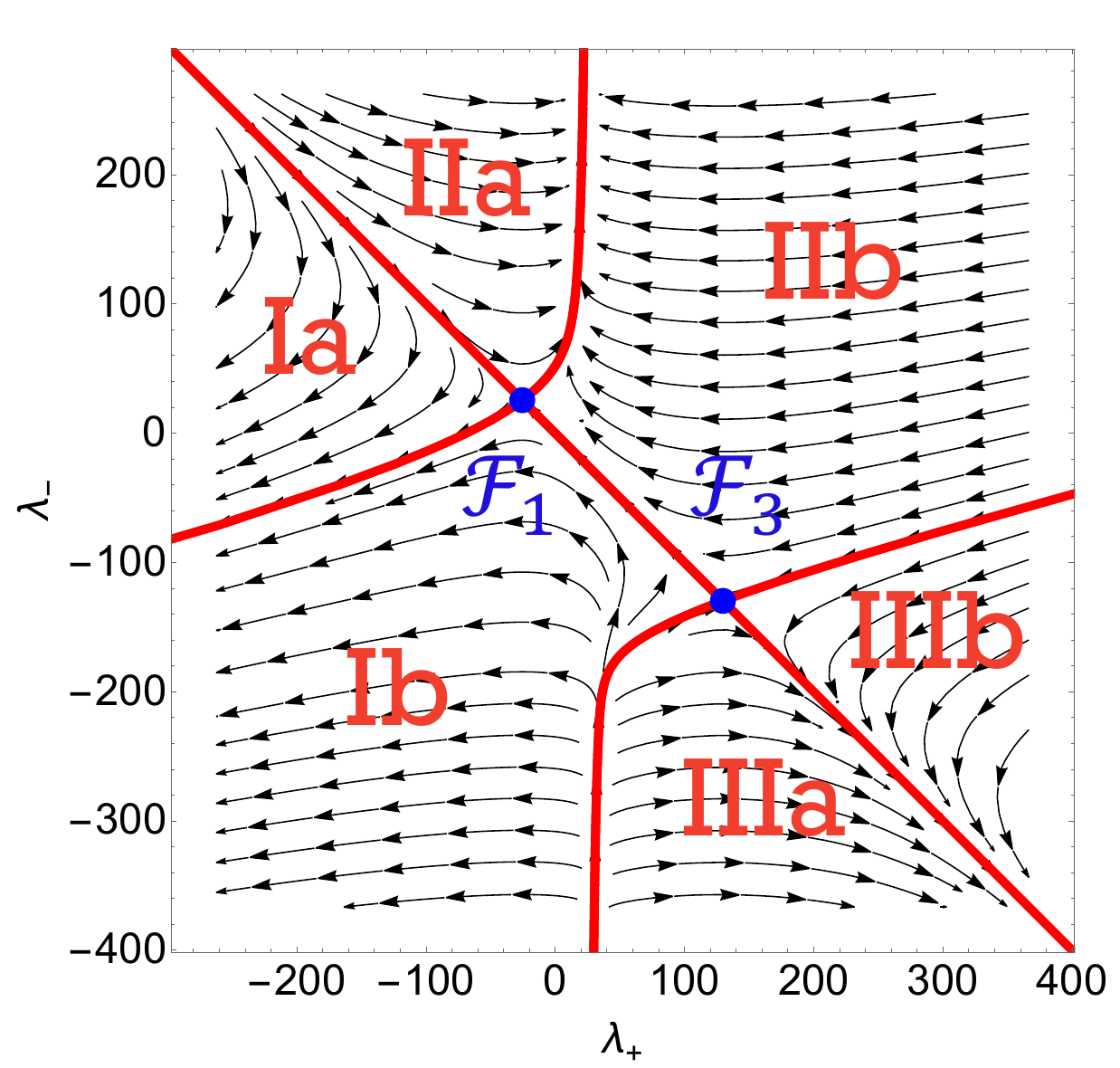}
    \caption{Phase diagram of the NJL theory subspace spanned by the 
$\left(\text{V} \pm \text{A} \right)$ channels at the Pauli-induced fixed point 
$\mathcal{C}$ ($\kappa^* = 3.82, \eta_\psi = -1$) for $\Nf = 1$ irreducible 
Dirac flavor. Arrows indicate flow towards the infrared. Compared to Figure 
\ref{fig:FPC_Nf2}, $\mathcal{F}_2$ and $\mathcal{F}_4$ lie at infinity. This is 
because the beta function \eqref{eq:LamMinusBeta} no longer depends on 
$\lambda_-$, with the simultaneous vanishing of the scaling term and fermionic 
loop contribution $\sim \lambda_-^2$.} 
    \label{fig:FPC_Nf1}
\end{figure}

We should note however, that the explicit violation of chiral symmetry by the 
Pauli term generates further four-fermion channels, e.g., an additional 
Gross-Neveu channel outside the NJL subspace. This occurs through the exchange 
of a Pauli-coupled photon between different flavors (in contrast to Figure 
\ref{fig:SingleExchange}). In 
Eqs.~\eqref{eq:LamPlusBeta}-\eqref{eq:LamMinusBeta}, we have discarded such 
contributions for simplicity. A Fierz-complete analysis of the RG relevance of 
such channels is beyond the scope of this work. Nevertheless, the 
structure of the resulting Fierz-complete equations remains similar to the 
chirally invariant subspace studied here: the inclusion of $n$ four-fermion 
channels can potentially entail $2^n$ fixed points in the corresponding 
coupling space $\lambda_i$. As long as one of these fixed points features 
properties of a shifted Gaussian fixed point similar to $\mathcal{F}_2$ for 
$\Nf>1$ in the present case, our main conclusions remain unaffected.

For completeness, let us mention that we have applied the same analysis 
to the compatibility of fixed point $\mathcal{B}$ discovered in 
\cite{Pauli_Gies_2020} with light fermions. As an approximation, we neglect 
additional terms in the beta functions due to the finite fermion mass $m$ and 
take into account such threshold effects only through the regulators in 
Eqs.~\eqref{eq:LamPlusBeta}-\eqref{eq:LamMinusBeta}. The most significant 
difference lies in the larger fermionic anomalous dimensions $\eta_\psi^* < -1$, 
which tends to reflect fixed points across the origin and reverse the direction 
of the RG flow. Once again, we observe a transcritical bifurcation between the 
now purely IR repulsive would-be Gaussian $\mathcal{F}_1$ and 
$\mathcal{F}_2$, but the latter is no longer responsible for avoiding heavy 
fermions. Instead, this role is taken up by the IR-stable $\mathcal{F}_4$, which 
lies at finite coupling values for $\Nf>1$.  

\section{Conclusion}
\label{sec:conclusion}

We have studied the renormalization flow of chirally invariant 
four-fermion operators when subject to a strong-coupling regime as provided by a 
recently discovered fixed point in QED including a non-Gaussion Pauli 
spin-field coupling. The flow of these fermionic operators is a crucial litmus 
test for the viability of an asymptotic safety scenario based on the 
non-Gaussian fixed point $\mathcal{C}$ as discussed in  
\cite{Pauli_Gies_2020,Pauli_https://doi.org/10.48550/arxiv.2210.11927}. This is 
because strong-coupling has the potential to drive chiral symmetry breaking in 
QED and generate a heavy electron mass incompatible with observation. 

In fact, Pauli-induced 
asymptotic safety at fixed point $\mathcal{C}$ demands that the fermionic 
anomalous dimension $\eta_\psi = -1$ renders four-fermion couplings 
perturbatively marginal, at least at the present level of truncation. At first 
glance, this  appears precarious as any propsective fixed points are then 
maximally susceptible to removal by the Pauli coupling term $\sim \kappa^4$ of 
the beta function.

However, the flow in the full chirally invariant plane spanned by the pointlike
four-fermion interactions known from NJL-type models captures the 
effects of single-photon exchange (Figure \ref{fig:SingleExchange}). The 
resulting coupled flow of the NJL-type couplings $\lambda_\pm$ 
exhibits transcritical bifurcations where the fixed points merely exchange 
stability, in stark contrast to the annihilation observed at strong 
minimal gauge 
coupling. As such, for $\Nf > 1$ irreducible Dirac flavors, there remains an 
infrared attractor at arbitrarily strong Pauli coupling, which 
prevents 
dynamical mass generation at a UV scale. This attractor is reminiscent 
of a shifted Gaussian fixed point. RG trajectories emanating from this fixed 
point are UV complete and do not introduce further physical parameters.
We observe that this scenario is not visible in a Fierz-incomplete 
chiral truncation based on a single NJL coupling. 

On the other hand, for $\Nf = 1$, the 
simultaneous vanishing of the scaling term and fermionic loop contribution 
$\sim 
\lambda_-^2$ conspire to prevent the existence of a fully attractive fixed 
point at finite coupling. While we do observe two fixed points, they do not 
satisfy all of our validity criteria and would come with further relevant 
directions, i.e., require a further physical parameter.  

Our work may be further extended to include a Fierz-complete basis including also the non-chiral four-fermion interactions. This would accommodate the Gross-Neveu 
channel generated by single-photon exchange between different flavors, resumming 
all ladder and crossed-ladder diagrams. The inclusion of further channels 
generically leads to an increase in the number of fixed points 
\cite{Gies:2003dp,Gehring:2015vja}. Our current scenario remains 
viable, if one of these fixed points remains fully IR attractive similar to a 
shifted Gaussian fixed point. While perfectly plausible, this remains to be 
confirmed.  

We have not fully considered the feedback of the four-fermion sector on the 
running of the gauge/Pauli couplings. While such a feedback on 
the minimal gauge coupling vanishes at the fermionic fixed points by virtue 
of the Ward-Takahashi identities \cite{Gies:2003dp}, a similar mechanism is 
not expected for the feedback on the flow of the Pauli coupling. But symmetry 
arguments ensure that such contributions be proportional to the mass $m$, which 
vanishes at fixed point $\mathcal{C}$ and can only affect the flow towards the 
IR. 

In summary, our findings provide further evidence for a  
scenario of an asymptotically safe UV completion of pure QED based on a 
non-Gaussian Pauli spin-field coupling. We identify a fixed-point collision 
with a subsequent exchange of stability properties as a crucial mechanism to 
avoid chiral symmtry breaking in the strong-coupling regime. This mechanism is, 
however, operative only for $\Nf>1$. 

\acknowledgments

We thank Jobst Ziebell for 
valuable discussions.  
This work has been funded by the Deutsche Forschungsgemeinschaft (DFG) under 
Grant Nos. 398579334 (Gi328/9-1) and 406116891 within the Research Training 
Group RTG 2522/1, as well as under Grant Nos.
392856280, 416607684, and 416611371 within the Research Unit FOR2783/2.


\bibliography{bibliography}
  
\end{document}